\documentclass[twocolumn,preprintnumbers,amsmath,amssymb,superscriptaddress]{revtex4}
\usepackage{epsf}
\usepackage{graphicx}
\usepackage{sidecap}
\usepackage{array}
\usepackage{amsmath}
\usepackage{amssymb}
\usepackage{dcolumn}
\usepackage{epstopdf}
\usepackage{bm}
\usepackage{amsmath}

\usepackage{color}

\def \beq {\begin{equation}}
\def \eeq {\end{equation}}
\begin{document}
\title{{Observation of topological nodal-loop state in RAs$_3$ (R = Ca, Sr)}}

\author{M.~Mofazzel~Hosen}
\affiliation {Department of Physics, University of Central Florida, Orlando, Florida 32816, USA}

\author{Baokai~Wang}
\affiliation {Department of Physics, Northeastern University, Boston, Massachusetts 02115, USA}

\author{Gyanendra Dhakal}
\affiliation {Department of Physics, University of Central Florida, Orlando, Florida 32816, USA}

\author{Klauss~Dimitri}
\affiliation {Department of Physics, University of Central Florida, Orlando, Florida 32816, USA}
\author{Firoza Kabir}
\affiliation {Department of Physics, University of Central Florida, Orlando, Florida 32816, USA}
\author{Christopher Sims}
\affiliation {Department of Physics, University of Central Florida, Orlando, Florida 32816, USA}
\author{Sabin Regmi}
\affiliation {Department of Physics, University of Central Florida, Orlando, Florida 32816, USA}
\author{Tomasz~Durakiewicz}
\affiliation {Institute of Physics, Maria Curie - Sk{\l}odowska University, 20-031 Lublin, Poland}

\author{Dariusz Kaczorowski}
\affiliation {Institute of Low Temperature and Structure Research, Polish Academy of Sciences,
	50-950 Wroc{\l}aw, Poland}

\author{Arun~Bansil}
\affiliation {Department of Physics, Northeastern University, Boston, Massachusetts 02115, USA}

\author{Madhab~Neupane}
\affiliation {Department of Physics, University of Central Florida, Orlando, Florida 32816, USA}

\date{\today}
\begin{abstract}
\noindent
{Topological nodal-line semimetals (NLSs) are unique materials, which harbor one-dimensional line nodes along with the so-called drumhead surface states arising from nearly dispersionless two-dimensional surface bands. However, a direct observation of these drumhead surface states in the currently realized NLSs has remained elusive. Here, by using high-resolution angle-resolved photoemission spectroscopy (ARPES) along with parallel first-principles calculations, we examine the topological characteristics of SrAs$_3$ and CaAs$_3$. SrAs$_3$ is found to show the presence of a topological nodal-loop, while CaAs$_3$ is found to lie near a topologically trivial phase. Our analysis reveals that the surface projections of the bulk nodal-points in SrAs$_3$ are connected by drumhead surface states. Notably, the topological states in SrAs$_3$ and CaAs$_3$ are well separated from other irrelevant bands in the vicinity of the Fermi level. These compounds thus provide a Ôhydrogen-likeÕ simple platform for developing an in-depth understanding of the quantum phase transitions of NLSs.}

\end{abstract}

\date{\today}
\maketitle
\noindent
\noindent
Experimental discoveries of non-trivial topological states in semimetals such as the Dirac \cite{Dai,Neupane_2, Cava1, NdSb}, Weyl \cite{TaAs_theory,TaAs_theory_1,Suyang_Science,Hong_Ding}, and nodal-line \cite{node_0, Neupane_5, Schoop, Ding} semimetals have greatly expanded the family of available topological materials beyond topological insulators \cite{Hasan, SCZhang, Xia, Hasan_review_2, Neupane_4, RMP}. The topological nodal-line semimetals (NLSs) are especially interesting as they host one-dimensional closed loops or line degeneracies in their electronic spectra. The density of states at the Fermi energy in an NLS is greater than that of a Dirac or Weyl semimetal, providing a more favorable condition for investigating exotic non-trivial phases and realistic materials platforms for developing applications. Note that, the NLSs are not robust against spin-orbit coupling or other perturbations and require crystal symmetries for their protection. To date, several structural classes of NLSs such as PbTaSe$_2$ \cite{PTS}, LaN \cite{LaN}, Cu$_3$PdN \cite{node_3}, and ZrSiS-type \cite{Schoop, Neupane_5,new_ZST, HfSiS, ZrGeM, ZrSiX, GdSbTe, ZrGeTe} materials have been reported with associated space group symmetries that protect the nodal-line state. However, the nodal-loop states in PbTaSe$_2$ \cite{PTS}, and Cu$_3$PdN \cite{node_3} lie in the vicinity of other metallic bands, LaN requires multiple symmetries for protection, while in the ZrSiX-type systems the topological states lie above the Fermi level. It is highly desirable, therefore, to find materials, which require minimum symmetry protections without the presence of other nearby bands that interfere in isolating topological features in the electronic spectrum.
	\noindent
	\begin{figure}
		\centering
		\includegraphics[width=8.9cm]{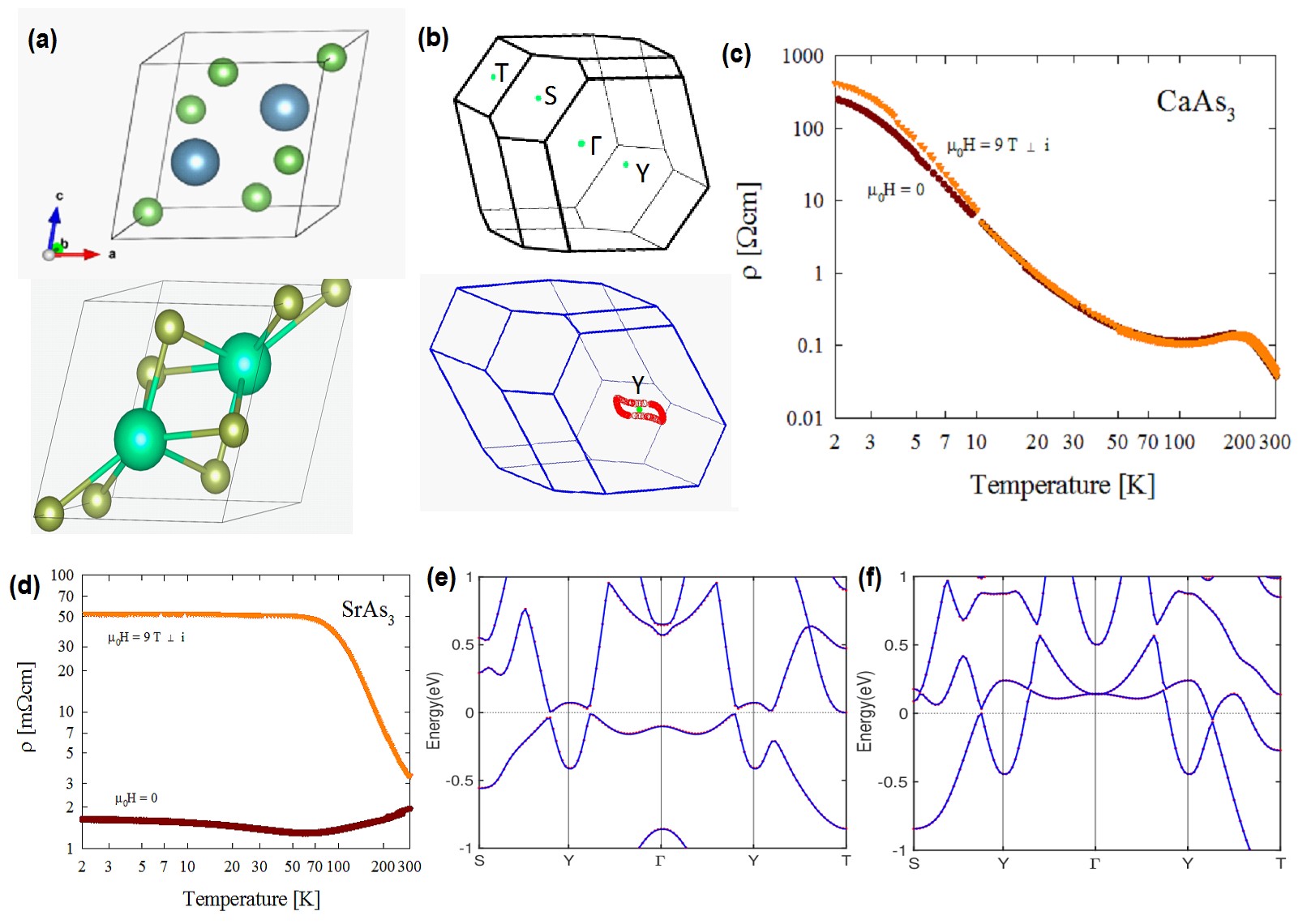}
		\caption{\textbf{ Crystal structure and sample characterization of RAs$_3$}. (a) Triclinic (up) and monoclinic (down) primitive unit cell. Purple (neon green) and green balls represent Ca(Sr) and As atoms, respectively. The center of inversion lies between the two neighboring Ca(Sr) atoms. (b) 3D Brillouin zone with the high symmetry points T, S, Y and $\Gamma$ (upper panel) are noted for SrAs$_3$. The nodal-line is located around the Y point for CaAs$_3$ (lower panel). 
(c),(d) Temperature dependent of the electrical resistivity (note double-logarithmic scales) of CaAs$_3$ and SrAs$_3$, respectively, measured in zero magnetic field and magnetic field of 9 T applied perpendicular to electric current. (e),(f) Bulk band structure along the high-symmetry points calculated without the inclusion of SOC for CaAs$_3$ and SrAs$_3$, respectively. Blue lines and dots correspond to the TB model and first-principles calculations, respectively.}
		\end{figure}

It has been recently shown that time-reversal symmetry (TRS) with a center of inversion symmetry (CIS) is sufficient, in principle, to protect a nodal-line state \cite{M1,M3,M2}. APn$_3$ (A = Ca, Sr, Ba, Eu; Pn = P, As) family of compounds has been identified as a potential materials class to host such a minimal symmetry protected NLS when SOC is excluded \cite{M3,M2}. Among these, CaP$_3$ and CaAs$_3$ are the only members of this series to have a triclinic crystal structure with space group \textit{P1}, known as the ``mother" of all space groups, whereas other members including SrAs$_3$ crystallized with higher symmetry structures characterized by space group \textit{C2/m}. Remarkably, in the \textit{P1} space group, CIS is the only crystalline symmetry that can protect the topological nodal-line states along with TRS \cite{M2}. Such a system can aptly work as the materials platform of an ideal nodal-loop system i.e as the primal ``hydrogen atom'' type nodal-loop. However, the experimental verification of this tempting conjecture has not been reported yet. 
Our studied material RAs$_3$ (R = Ca,Sr) could provide not only the nodal-loop state but also the topological surface states or drumhead surface states in momentum space connecting nodal points. RAs$_3$ thus appears to be a system with an enhanced topological density of states, paving the road for the potential discovery of more exotic states. 

Here, we report the observation of a topological nodal-loop state in the monoclinic system SrAs$_3$ and trivial state in triclinic system CaAs$_3$. Utilizing angle-resolved photoemission spectroscopy (ARPES), we systematically study the detailed electronic structure of these materials. Our ARPES data and first-principles calculations reveal the presence of a topological nodal-loop state around the center (Y) of the Brillouin zone (BZ) in SrAs$_3$. Furthermore, we observe a drumhead surface state connecting the nodal-point projection along the Y direction. Our experimental data are corroborated by our first-principles calculations. Interestingly, our calculations suggest that CaAs$_3$ undergoes a topological phase transition from TNL to TI phase with surface states which are practically flat when SOC is turned on (also see Ref. \cite{M3,M2}). Furthermore, our experiments reveal that the Fermi surface of CaAs$_3$ is formed by a single isolated band. Therefore, our study could open up a new platform for studying the interplay between various topological phases.\\~\\

Single crystals of RAs$_3$ were grown by Sn-self flux technique (see Supplementary Information (SI) for details). The first-principles calculations were performed using both the density functional theory (DFT) and tight binding (TB) methods (see SI).
Synchrotron-based ARPES measurements were performed at the SIS-HRPES end-station at the Swiss Light Source (SLS) equipped with Scienta R4000, Advanced Light Source (ALS) beamline 10.0.1 equipped with Scienta R4000 and ALS beamline 4.0.3 equipped with R8000 hemispherical electron analyzers. The energy resolution was set to be better than 20 meV, and the angular resolution was set to be better than 0.2$^\circ$. Samples were cleaved in situ and measured at 10 - 25 K in a vacuum better than 10$^{-10}$ torr. The (Ca/Sr)As$_3$ specimens were found to be very stable and did not show any sign of degradation for the typical measurement period of 20 hours.\\~\\

The triclinic crystal structure of CaAs$_3$ is shown in the upper panel of Fig. 1(a). The center of inversion lies midway between the neighboring Ca atoms. The crystal structure of SrAs$_3$ has higher symmetry compared to that of CaAs$_3$ (see Fig. 1(a)), hence, SrAs$_3$ crystallizes in a simple monoclinic structure with space group \textit{C2/m}. Therefore, in addition to center of inversion symmetry, SrAs$_3$ has \textit{C2} rotational symmetry. The center of inversion symmetry lies midway between two Sr atoms and the two fold rotational symmetry can be readily observed from the primitive monoclinic unit cell as shown in Fig. 1(a) (lower panel). The corresponding bulk Brillouin zone with high symmetry points is shown in the upper panel of Fig. 1(b). An important point to note that the projections of Y and $\Gamma$ points on the (010) plane are located at the same point of the BZ. The lower panel of Fig. 1(b) demonstrates the location of a nodal-loop centered around the Y point while the spin-orbit coupling (SOC) effect is excluded; here a little deviation from the S-Y-T plane is observed for CaAs$_3$ (note that it is in  the plane for SrAs$_3$).

The electrical transport measurements indicate a semiconducting character of CaAs$_3$ (see Fig. 1(c)), in agreement with the previous reports \cite{struc2, rho}. At room temperature, the resistivity is about 37 m$\Omega$cm, and with decreasing temperature it increases non-monotonically, initially in a semimetallic manner, passing through a smeared shallow maximum near 200 K, but then rising sharply below 15 K. The resistivity measured at 2 K is about 260 $\Omega$cm, which is a value nearly four orders of magnitude larger than that at 300 K. The overall shape of $\rho (T)$ as well as the values of the resistivity are very similar to those reported in the literature \cite{struc2, rho}. As can be inferred from Fig. 1(c), an external magnetic field of 9 T, applied perpendicular to the electric current, hardly affects $\rho(T)$ of CaAs$_3$ above 10 K, yet brings about a more rapid rise of the resistivity at lower temperatures (the resistivity achieved at 2 K is by 60\% larger than that measured in zero field). The latter feature can be attributed to small increase in the value of the semiconducting energy gap or/and some reduction in the mobility of dominant charge carriers, both effects being driven by magnetic field. 

The electrical transport behavior of SrAs$_3$ is presented in Fig. 1(d). In zero magnetic field, the compound exhibits semimetallic properties with weakly temperature-dependent resistivity of about 1.5 mWcm and a shallow minimum in $\rho (T)$ near 60 K, in concert with the literature data \cite{rho,Transport}. In a magnetic field of 9 T, applied perpendicular to the electric current, the resistivity of SrAs$_3$ notably changes. In the region from room temperature down to about 70 K, the compound shows semiconducting-like behavior, while at lower temperatures, a plateau in $\rho (T)$ is observed, at which the resistivity amounts to $\sim$50 mWcm, i.e. it is 3000\% larger than the magnitude in zero field. Such a distinct influence of magnetic field on the electrical transport in SrAs$_3$ can be attributed to field-induced changes in mobilities and concentrations of electron and hole carriers in a two-band material. Similar picture was invoked before to explain unusual galvanomagnetic properties of SrAs$_3$, like first-order longitudinal Hall effect and magnetoresistivity in Hall geometry \cite{Gal}.

Figures 1(e) and 1(f) show the bulk electronic band structure of CaAs$_3$ and SrAs$_3$, respectively, calculated along the various high symmetry directions using tight-binding (lines) and first-principles (dots) techniques without considering spin-orbit coupling (SOC) effects. Analyzing the calculations of both materials without SOC, one finds a nodal-loop around the Y point of BZ, which is located in the vicinity of the chemical potential. An important point to note that the bands are fully gapped as they diverge from the Y points in both directions. The small gap in CaAs$_3$ is due to the fact that the nodal points lie slightly away from the high symmetry points. The inclusion of SOC results in opening a negligible gap in SrAs$_3$ and approximately 40 meV gap in CaAs$_3$ along the Y-$\Gamma$ direction (see SI). The observed instability of the nodal-loop phase against the fully-gapped topological phase is in concert with the experimental electrical resistivity data of CaAs$_3$, where a crossover from semimetallic to low-temperature insulating behavior occurs. However, the insulating character is sufficiently weak enough to neglect in SrAs$_3$ for both the transport measurements (see Fig. 1(d)) and first-principles calculations (see SI). An important point to note that the exclusion of SOC to observe nodal-line or loop states is a well-known prevalent technique that has played a significant role in realizing previously reported nodal-line semimetals such as LaN \cite{LaN}, Cu$_3$(Pd,Zn)N \cite{CPN, node_3}, ZrSiX-type materials \cite{Schoop, Neupane_5, ZrSiX}, etc.\\~\\

In order to determine the nature of charge carriers and to unveil the Fermi surface evolution with the binding energy, we present the Fermi surface and constant energy contour plots in Figure 2. The hexagonal Fermi surface of SrAs$_3$ is observed with 55 eV incident photon energy in Fig. 2(a).  At the center, we clearly observe a circular pocket, which is a result of the surface arc-like state near the Fermi level, namely the drumhead surface state. Furthermore, we observe six petal-like pockets resembling a flower-like shape. Moving towards higher binding energy (170 meV), we observe that the circular pocket almost disappears and the six petals begin to overlap each other creating a complex feature. The oval shape at the corner also evolves into a small point-like shape. At around 600 meV below the chemical potential the oval shape and the circular pocket at zone center completely disappear indicating the electron-like nature of the bands around these points. However, the six-petal-flower shape evolves into a complex flying bat like feature and confirms the hole-like nature of these bands. From the bulk band calculations, one can easily speculate that the six petals will form a bigger nodal ring around the drumhead surface state. However, the corner of the Brillouin zone is not well resolved at this photon energy. Therefore, we mapped the Fermi surface at a higher incident energy (100 eV) at the SLS beamline, which further confirms the hexagonal nature of the Brillouin zone (BZ) (see Fig. 3(a) and SI). Figure 2(b) (left) shows the experimental Fermi surface map of CaAs$_3$ within a wide momentum window. Unlike SrAs$_3$, we do not observe the electron-like pockets and flower petal shape at the corner and center of the BZ. Each of the hexagons observed represents an individual BZ of CaAs$_3$. In order to figure out the evolution of the Fermi contour, we present the constant energy contour plots in Fig. 2(b)(right) and in SI. In these figures, one can clearly see the distorted hexagonal shape of the BZ and hole-like nature of the carrier, that is perfectly reproduced by our calculations (see SI).\\~\\

\begin{figure}
	\centering
	\includegraphics[width = 8.9cm]{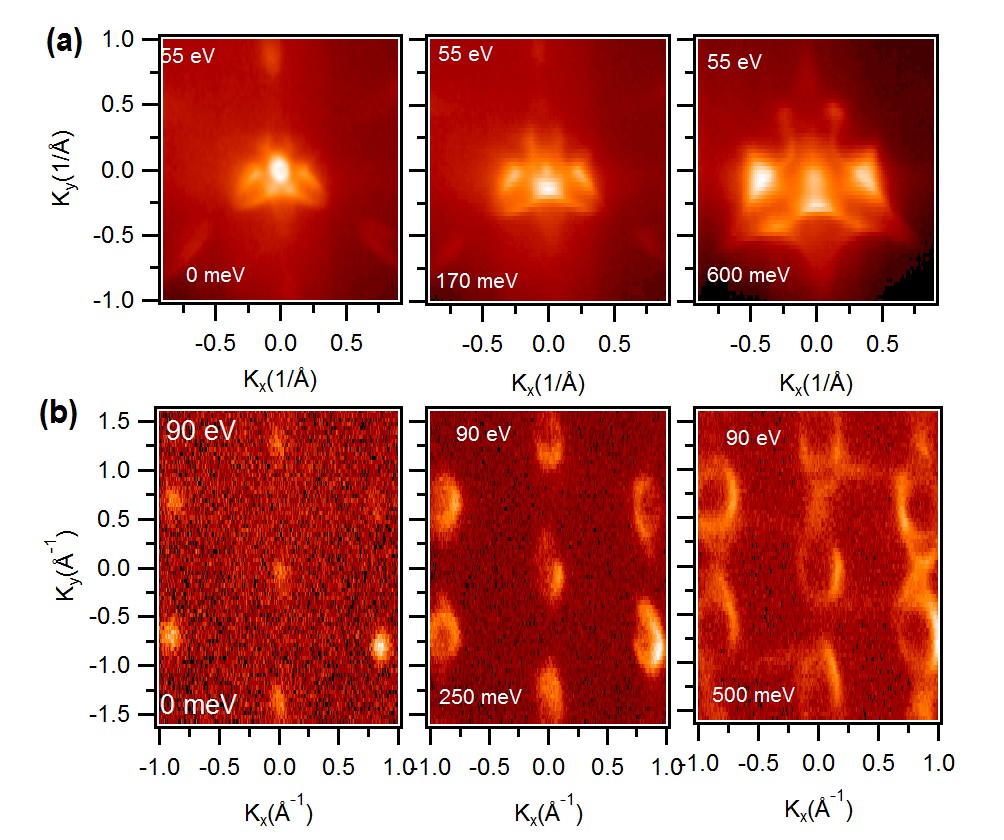}
	\caption{{Fermi surfaces and constant energy contours of RAs$_3$}  (a) Results for SrAs$_3$ obtained at the ALS beamline 10.0.1 using a photon energy of 55 eV. (b) Results for CaAs$_3$. Each of the distorted hexagons represents a separate Brillouin zone where the b-axis is larger than the a-axis. These measurements were performed at the HRPES end-station of SLS beamline at a temperature of 18 K using an incident photon energy of 90 eV. The binding energies are marked in the plots. }
\end{figure}

In order to determine the nature of the electronic bands associated with the nodal-loop near the Fermi level, the photon energy dependent energy-momentum dispersion maps were measured (see Figs. 3 and 4, and SI). Figure 3(a) shows the Fermi surface map measured at 100 eV photon energy. The white dashed lines represent the cut directions for the energy-momentum dispersion measurements. Figure 3(b) shows the photon energy dependent dispersion maps along the cut 1 direction of SrAs$_3$. Here, we observe the 2D Fermi surface states which correspond to the drumhead surface states at the $\Gamma$-point at all the photon energies. The bulk bands below the surface states are not well resolved at the low photon energies, therefore we plot the dispersion map at 100 eV (see Fig. 3(c)). Here, one can clearly observe the bulk bands, providing an explanation for the flying bat-like shape of the BZ at higher binding energies. Furthermore, the Dirac point of the nodal loop and the arc along the Y-$\Gamma$ direction meet in the vicinity of the Fermi level. Most importantly, the arc-like state does not show any notable dichotomy with photon energies, hence, we conclude that it is surface originated (also see Fig. 3(d) for guide to eyes and SI). This further confirms our observation of the drumhead surface states and the nodal loop which is in agreement with our first-principles calculations (see Fig. 1(f) and Ref. \cite{M3,M2}). Next, we represent the dispersion maps along the six electron pockets observed at the corner in Fig. 3(e) (cut 2 direction). A massive Dirac like state is observed with $\sim$0.3 eV gap size.\\
\noindent
\begin{figure*}
	\centering
	\includegraphics[width=18cm]{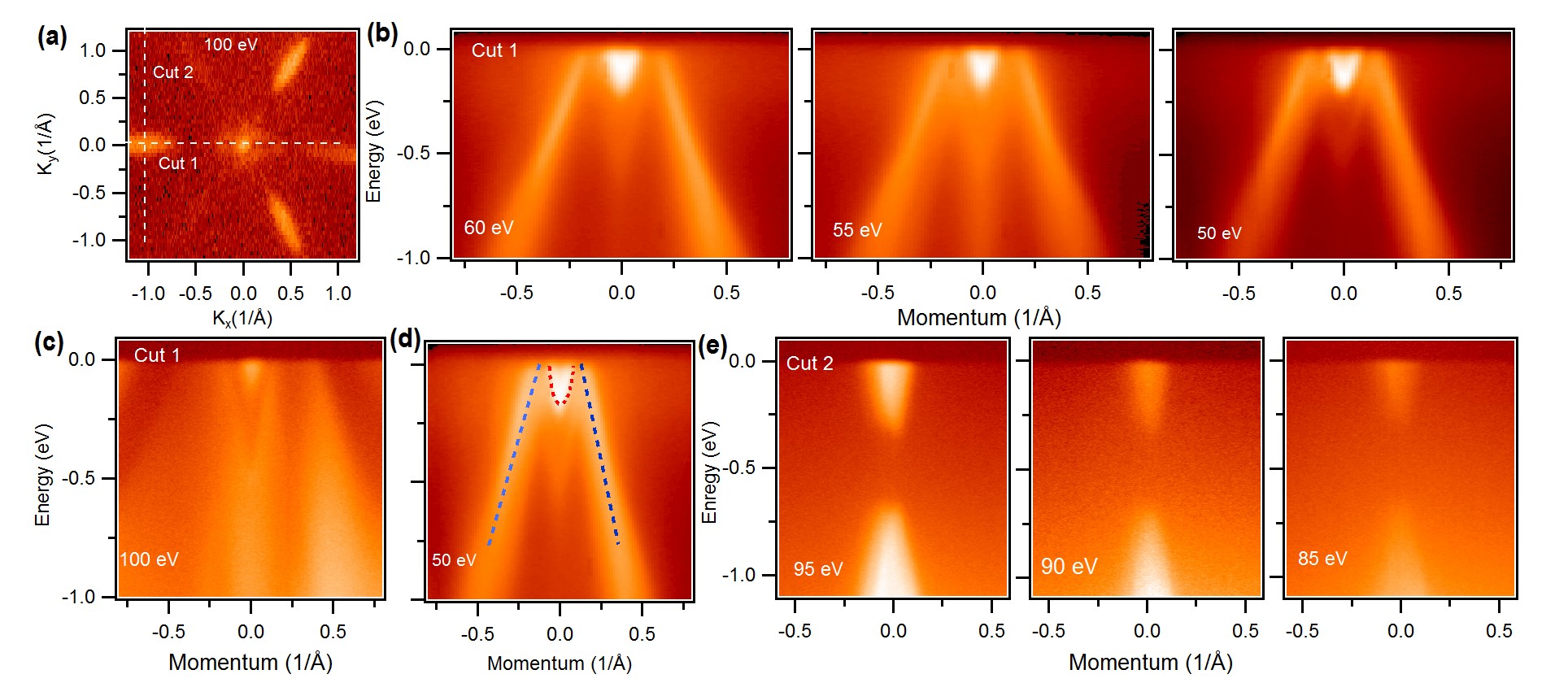}
	\caption{{Observation of nodal-loop state in SrAs$_3$.} (a) Fermi surface map at a photon energy of 100 eV. (b) Photon energy dependent dispersion maps along the cut 1 direction. (c) Dispersion map measured at 100 eV photon energy. (d) Dispersion map with guide to the eyes. The red dotted arc shows the drumhead surface state. (e) Dispersion map along the center of the electron-like hole pocket. All measurements were performed at the ALS beamline 10.0.1 and SLS at a temperature of 18 K. The photon energies are marked in the plot.}
\end{figure*}

Figure 4 represents the dispersion maps along the various direction of CaAs$_3$ shown in Fig. 4(a). We use several photon energies for probing different values of the perpendicular components of the crystal momentum. From the results presented in Fig 4(b) (see also SI), it is clear that only a single band appears in the vicinity of the chemical potential without any interference from irrelevant bands. Our calculations suggest that the surface states lie within the upper part of the band, which is located slightly below the chemical potential. Such a naturally tuned clean system in the vicinity of the chemical potential is very crucial for transport behavior as well as for applications. To understand the nature of the bands along this high-symmetry direction in the (010) plane, we carried out the band-dispersion calculations without (see Fig. 4(c)) and with (see SI) the inclusion of SOC. The nodal loop (without SOC) and the surface state (with SOC) are found around the Y point along the $k_x$ momentum plane.  Most interestingly, the projections of the nodal points in the $k_x$ direction are connected by the surface states. However, the inclusion of SOC opens a 40 meV gap along this direction and the system undergoes a topological phase transition from NLS to TI phase. The observation of the small gap in Fig. 4(b) supports the gap opening phenomena. Closely looking at the surface state of CaAs$_3$, we show a zoomed-in view of the experimental dispersion map, and results of calculations performed near the chemical potential by including SOC (see Figs. 4(d)-(e)). Interestingly, in Fig. 4(d) we do not see the surface state within the top part near the Fermi level but a finite gap is observed. Therefore, we conclude that our experimentally observed state in CaAs$_3$ is topologically trivial in nature. With no other bands near the Fermi level, CaAs$_3$ thus provides a unique opportunity to see the evolution of NLS phase to TI phase by small doping. Fig. 4(e) shows the calculated dispersion map near the Fermi level where one can see the almost flat surface state connecting the bulk bands. Fig. 4(f) represents the measured dispersion map along the k$_y$ directions which clearly supports our previous observations. Here, we observe that the band is almost flattened in the k$_y$ direction while we find a sharp dispersion along the k$_x$ direction. This could further provide a tuning knob to study more exciting exotic states. \\~\\

Now we turn to a discussion of the possible nature of the major observed phenomena. Our study reveals the presence of a hexagonal shaped Fermi surface and the topological nodal loop in RAs$_3$. Based on the crystallographic data, the observed topological states in CaAs$_3$ are shown to be protected by CIS only, and their presence is reproduced in calculations without taking the SOC effect into account. The lack of other symmetries in a triclinic (P1) lattice, and the absence of any other bands in the vicinity of the Fermi energy lead to the simplest form of a topological nodal-loop in this material when SOC is neglected. On the contrary, the inclusion of SOC in CaAs$_3$ leads to a gap-opening where an energetically almost flat surface state appears. The length of the connecting arc along the \textit{k$_x$} is estimated to be about 0.08 \textup{\AA}$^{-1}$, and it is found to be much longer along the \textit{k$_y$} direction at about 0.30 \textup{\AA}$^{-1}$ (SI). The experimental transport data comply with the latter result showing that the bulk resistivity of the compound at 2 K is as large as 260 $\Omega$-cm. However, our experimental data shows a clear gap near the Fermi level and we conclude CaAs$_3$ to be a topologically trivial material. On the other hand, SrAs$_3$ has two-fold rotational symmetry along with the CIS. We emphasize that we observe the nodal loop and the Fermi surface arc connecting the projection of the nodal-point along the \textit{$k_x$} momentum plane. Importantly, the surface states in SrAs$_3$ form a drumhead surface state around the $\Gamma$ point and the observation of such an in-plane drumhead surface state is the first of its kind. Our observation is also supported by recently reported Shubnikov-de Haas oscillation measurements in SrAs$_3$ \cite{Transport}. Therefore, by appropriate isoelectronic doping in CaAs$_3$ by Sr (Sr$_{1-x}$Ca$_x$As$_3$), we expect to observe the transition from NLS phase to TI. The topological surface states are expected to be in the vicinity of the Fermi level in Sr$_{1-x}$Ca$_x$As$_3$. 
Our experimental and theoretical results support our conclusion that RAs$_3$ would provide an excellent materials platform for comprehensive studies of the topological aspects of band theory and the interplay between the topological states without the interference from other metallic bands. Furthermore, our demonstration of the nodal-loop state, drumhead states in SrAs$_3$ and the nodal loop in a minimal symmetry protected naturally tuned system should provide an ideal platform to study topological physics.\\~\\
 \begin{figure}
 	\centering
 	\includegraphics[width=9cm]{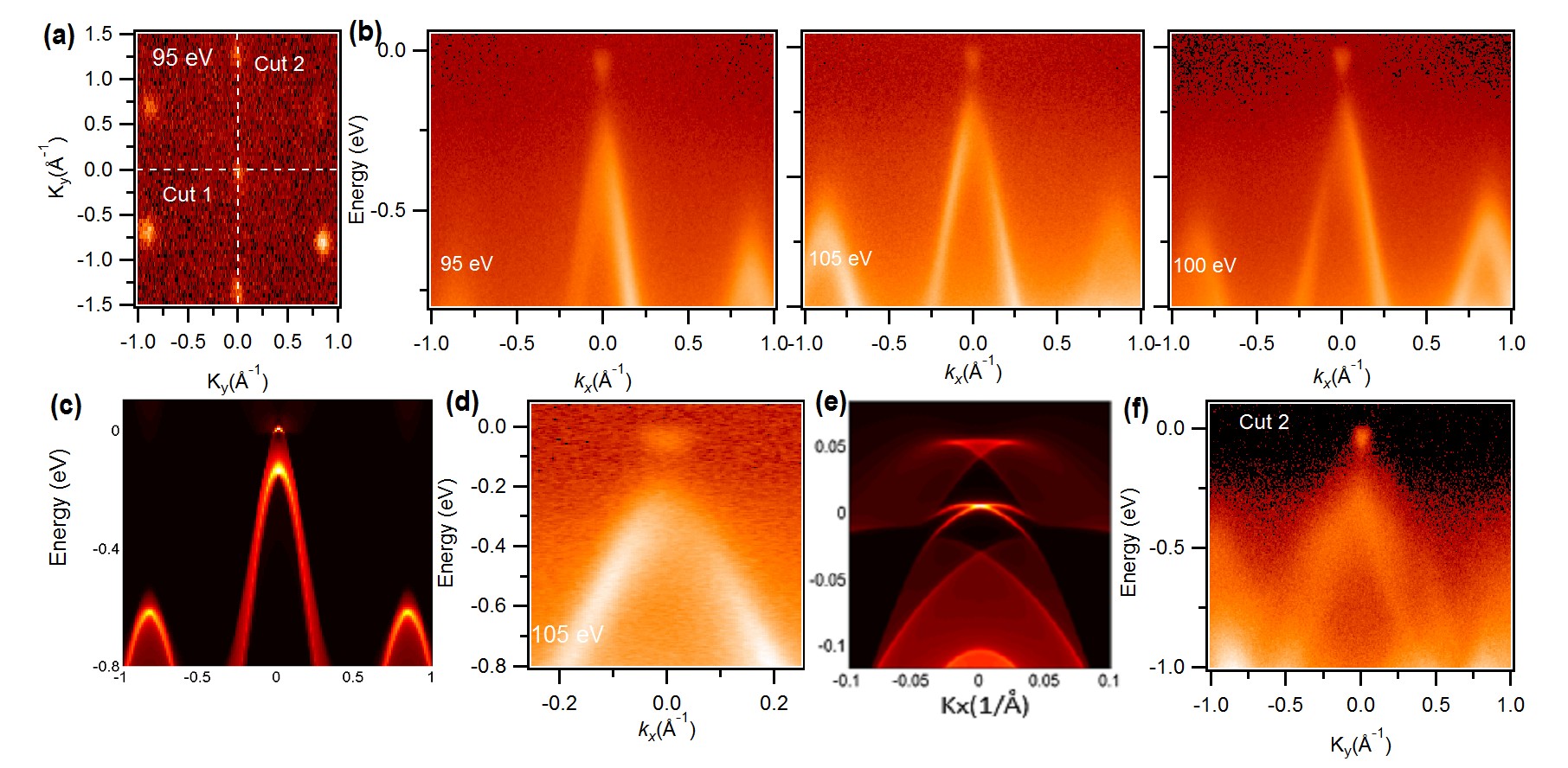}
 	\caption	{{Dispersion maps along the high symmetry directions in CaAs$_3$.} {(a) Fermi surface map at a photon energy of 95 eV. White dashed line guides the energy-momentum dispersion measurement directions. (b) Photon energy dependent dispersion maps along the cut 1 direction shown in Fig. 4(a). (c) Calculated dispersion map without the inclusion of SOC around the Y point of BZ. At the (010) surface, Y and $\Gamma$ are projected at the same point. (d) Zoomed in plot of 105 eV dispersion map. (e) Calculated zoomed-in plot near the Fermi level. (f) Measured dispersion map along the cut 2 direction. }}
 \end{figure}

In conclusion, we have performed systematic ARPES measurements along with parallel first-principles calculations on SrAs$_3$ and CaAs$_3$ in order to delineate the topological characteristics of these compounds. Our study reveals the presence of topological-nodal-loop state and drumhead surface states in SrAs$_3$, while CaAs$_3$ is found to be in the vicinity of a trivial phase. Our analysis indicates that these compounds could undergo topological phase transitions between the nodal-line-semimetal and topological-insulator phases with nearly flat surface states with appropriate tuning the spin orbit coupling and provide a robust platform for investigating the interplay of these quantum phases.

\noindent
\textbf{Acknowledgment}\\
\noindent
M.N.\ is supported by the Air Force Office of Scientific Research under Award No. FA9550-17-1-0415 and the startup fund from UCF. T.D. is supported by NSF IR/D program. D.K. is supported by the National Science Centre (Poland) under research grant 2015/18/A/ST3/00057. The work at Northeastern University was supported by the US Department of Energy (DOE), Office of Science, Basic Energy Sciences grant number DE-FG02-07ER46352, and benefited from Northeastern University's Advanced Scientific Computation Center (ASCC) and the NERSC supercomputing center through DOE grant number DE-AC02-05CH11231. We thank Sung-Kwan Mo and Jonathan Denlinger for beamline assistance at the LBNL. We also thank {Nicholas Clark Plumb} for beamline assistance at the SLS, PSI.\\~\\



\vspace{0.5cm}
\def\bibsection{

\bigskip

\noindent
M.M.H. and B.W. contributed equally to this work. 

Correspondence and requests for materials should be addressed to M.N. (Email: Madhab.Neupane@ucf.edu).

\noindent

\end{document}